\documentclass[a4paper,10pt,twoside]{cpc-hepnp}

\usepackage{multicol}
\usepackage{graphicx}
\usepackage{booktabs}
\usepackage{amssymb,bm,mathrsfs,bbm,amscd}
\usepackage[tbtags]{amsmath}
\usepackage{lastpage}

\usepackage{units}
\usepackage{amsfonts}
\usepackage{dsfont}

\newcommand{\conjg}[1]{\ensuremath{\hspace{1pt}\overline{\hspace{-1pt}#1\hspace{-1pt}}}\hspace{1pt}}

\def\Slash#1{\setbox0=\hbox{$#1$} 
\dimen0=\wd0 
\setbox1=\hbox{/} \dimen1=\wd1 
\ifdim\dimen0>\dimen1 
\rlap{\hbox to \dimen0{\hfil/\hfil}} 
#1 
\else 
\rlap{\hbox to \dimen1{\hfil$#1$\hfil}} 
/ 
\fi}

\begin{document}

\fancyhead[co]{\footnotesize R.\ Alkofer et al: Hadron properties from QCD
bound-state equations }

\title{Hadron properties from QCD
bound-state equations:\\ A status report}

\author{Reinhard Alkofer$^1$\email{reinhard.alkofer@uni-graz.at}
\quad Gernot Eichmann$^{1,2}$,
\quad Andreas Krassnigg$^1$,
\quad Diana Nicmorus$^1$
}
\maketitle

\address{
1~Institut f\"ur Physik, Karl-Franzens-Universit\"at Graz, 8010 Graz, Austria\\
2~Institute for Nuclear Physics, Darmstadt University of Technology, 64289
Darmstadt, Germany
}

\begin{abstract}
Employing an approach based on the Green functions of Landau-gauge QCD, some
selected results from a calculation of meson and baryon properties are
presented. A rainbow-ladder truncation to the quark Dyson-Schwinger
equation is used to arrive at a unified description of mesons and baryons by
solving Bethe-Salpeter and covariant Faddeev equations, respectively.
\end{abstract}



\begin{multicols}{2}


\section{Motivation:\\ Why Functional Approaches to QCD?}

The central aim of the studies reported here is to develop a  QCD--based
description of the structure of hadrons in  terms of quarks and gluons.
Theoretical issues such as confinement, dynamical breaking of chiral symmetry
and the formation of relativistic bound states can be understood and related to
the properties of QCD's Green functions.
Significant progress within functional
approaches has recently been achieved especially in Landau gauge, see {\it e.g.} refs.\
\cite{Fischer:2009tn,Fischer:2008uz,Alkofer:2008tt,Alkofer:2004it}.
Hereby it should be noted that different pictures of confinement are not
necessarily mutually exclusive  but may describe different facets of this
phenomonen, see \cite{Alkofer:2006fu} and references therein. The advantage of
employing Green function methods is given by the fact that these Green functions
provide the input for QCD's bound state equations which, in turn, can be used to
calculate hadron properties. The Green functions of elementary fields are
determined by Dyson-Schwinger equations, Exact Renormalization Group equations,
and/or lattice calculations. It should be noted that dynamical breaking of
chiral symmetry leads to the generation of quark masses and scalar quark-gluon
interactions \cite{Alkofer:2008tt,Alkofer:2006gz}.

Any numerical solution of QCD's bound state equations, {\it i.e.}~the
Bethe-Salpeter equation for mesons and the Faddeev equation for baryons,
requires Green functions of quarks and gluons as input, which in turn
necessitates a truncation  of the corresponding functional equations.
The results which will be presented in this
status report  build upon a rainbow-ladder truncation,
{\it i.e.}~a dressed-gluon exchange, for mesons and baryons in a unified
approach~\cite{Eichmann:2009zx,Nicmorus:2008vb,Eichmann:2008kk,Nicmorus:2008eh,Eichmann:2008ef}.
A symmetry-preserving extension to more realistic kernels is on its way for mesons, see
\cite{Fischer:2009jm} and references therein.
For baryons as three-quark states, the covariant Faddeev equation in rainbow-ladder truncation
has been solved only recently~\cite{Eichmann:2009qa}.


            \begin{figure*}[tbp]
            \begin{center}
            \includegraphics[scale=0.6]{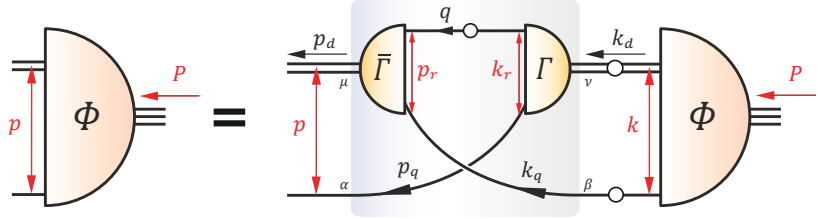}
            \caption[Quark-Diquark BSE]{The quark-diquark BSE in pictorial form. } \label{fig:qdqbse}
            \end{center}
            \end{figure*}

\section{Bound state equations and\\ rainbow-ladder truncation}

            Hadrons, being bound states of quarks, appear as poles in the $q\bar{q}$ and $qqq$ scattering matrices.
            Hadron properties can be extracted upon
            solving bound-state equations which are valid at these poles and need the elementary QCD Green functions as input:
            \begin{equation}
                \Psi = \widetilde{K}_\text{(n)}\,\Psi\, . 
            \end{equation}
            Here $\Psi$ is the bound-state amplitude defined on the hadron's mass shell,
            and $\widetilde{K}_\text{(2)} = K_\text{(2)} SS$, $\widetilde{K}_\text{(3)} = K_\text{(3)} SSS$ are the renormalization-group invariant products
            of the respective kernel $K_\text{(n)}$ with two or three dressed quark propagators.

            These fully Poincar\'e-covariant bound-state equations provide a tool to calculate experimentally accessible hadron observables, 
            {\it e.g.\/} meson and baryon mass spectra, decay constants, scattering processes, and electromagnetic properties 
            such as form factors, magnetic moments and charge radii (see {\it e.g.}~Refs.~\cite{Krassnigg:2009zh,Nicmorus:2008vb} 
            and references therein). To proceed with the numerical solution of a covariant bound-state equation like
            the Bethe-Salpeter or the Faddeev equation, one needs to specify all ingredients: the interaction kernels 
            and the dressed quark propagator which appear in these integral equations. These Green functions can be determined either 
	    from functional equations and/or lattice calculations. In addition, they are related via Slavnov-Taylor and Ward-Takahashi 
	    identities. With respect to the pion, the would-be Goldston boson of chiral symmetry, the 
            axial-vector Ward-Takahashi identity plays a special role.
            If it is satisfied by the interaction kernels in related equations, this correctly implements
            chiral symmetry and its dynamical breaking, leading {\it e.g.}~to a generalized Gell-Mann--Oakes--Renner relation
            which is valid for all pseudoscalar mesons and all current-quark masses~\cite{Maris:1997hd,Holl:2004fr}.
            In particular, the pion is massless in the chiral limit independent of the details of the interaction.
            
            Although the structure of the kernel is restricted by the above mentioned identities, to construct 
	     a symmetry-preserving truncation of the bound-state
	    equations is a non-trivial task. The rainbow-ladder truncation provides such a symmetry-preserving scheme. Instead of interaction kernels
	    being functions of the 
	    quark and gluon momenta one needs a functional form for the interaction as a function
            of the gluon momentum only. The non-perturbative dressing of the gluon propagator and the quark-gluon
	        vertex are ``absorbed'' into an effective coupling for which we adopt a widely-employed ansatz~\cite{Maris:1999nt,Eichmann:2008ae}.
            It reproduces the logarithmic decrease of QCD's one-loop perturbative running coupling at high momenta, and its
            infrared contribution is parametrized by an infrared scale and a dimensionless parameter (for details, see \cite{Eichmann:2008ae}).
            Especially, it yields the non-pertur\-bative enhancement at intermediate gluon
	        momenta necessary to generate dynamical chiral symmetry breaking, and hence a constituent-quark mass scale.
            We emphasize here that the input of the Faddeev equation is completely specified at the beginning
            with all parameters already fixed from a few meson properties. In addition, we want to note that the resulting quark propagator is
	    in very good agreement with corresponding results of fully coupled Dyson-Schwinger, resp., Functional Renormalization Group equations 
	    and of lattice calculations.




            \begin{figure*}[tbp]
                    \begin{center}
                    \includegraphics[scale=0.45]{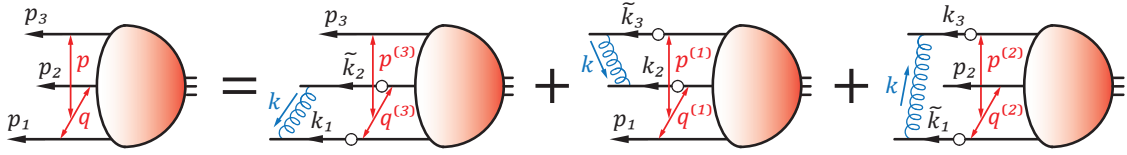}
                    \caption{Faddeev equation \eqref{faddeeveq} in rainbow-ladder truncation.}\label{fig:faddeev}
                    \end{center}
            \end{figure*}

\section{Baryons: Quark-diquark model} \label{sec:qdq}

            As the amplitudes resulting from a fully relativistic Faddeev equation are of high complexity it is natural to look first for an approximation keeping 
	    Poincar\'e invariance and the relation to QCD Green functions. A well-studied approximation of this type is based on using diquark correlations as effective
	    degrees of freedom.

            The motivation for studying diquark correlations has been the observation of a strong attraction in the $SU(3)_C$ antitriplet $qq$ channel,
            {\it e.g.}~in lattice \cite{Hess:1998sd,Wetzorke:2000ez,Orginos:2005vr,Liu:2006zi,Alexandrou:2006cq,Babich:2007ah} and Bethe-Salpeter \cite{Maris:2002yu,Maris:2004bp} studies.
            Such an attraction has also been proposed to explain missing exotic states in the hadron spectrum and the masses of light scalar mesons \cite{Anselmino:1992vg,Jaffe:2004ph}.
            Further support for the diquark concept has been provided by a study of diquark confinement in Coulomb-gauge QCD \cite{Alkofer:2005ug}.

            These arguments lead to the assumption that correlations between two quarks provide the dominant attraction not only in meson but also in baryon channels.
            Consequently, the quark-diquark model traces the nucleon's binding to the intrinsic formation of colored scalar- and axialvector diquarks.
            It treats such two-quark correlations as a separable pole sum in the $qq$ scattering matrix
            which leads to a description of baryons as bound states of effective quarks and diquarks.
            In this way the covariant Faddeev equation
            (see Section~{\ref{sec:faddeev}})
            is simplified to a two-body bound-state equation (see Fig.~\ref{fig:qdqbse})
            while full Poincar\'e covariance is maintained.

            Nucleon and $\Delta$ properties have been studied in a quark-diquark model with parametrized ingredients, see
	    {\it e.g.}~\cite{Hellstern:1997pg,Oettel:2000jj,Alkofer:2004yf,Holl:2005zi} and references therein.
            The approach was subsequently extended to determine the dynamics of $0^+$ and $1^+$ diquarks from their underlying quark and gluon constituents
            \cite{Eichmann:2007nn,Eichmann:2008ef,Nicmorus:2008vb,Eichmann:2008kk,Nicmorus:2008eh} where
            parametrizations for the diquark amplitudes were removed and replaced by solutions of the corresponding diquark Bethe-Salpeter equations.

            The identification of colored diquarks as poles in the $qq$ scattering matrix is possible within a rainbow-ladder truncation:
            one obtains timelike $0^+$, $1^+$,~$\ldots$~diquark poles. The corresponding mass scales play an important role in the description of light baryons.
            These poles, however, correspond to unphysical asymptotic states and disappear from the spectrum when going beyond rainbow-ladder.
            The obtained rainbow-ladder diquark masses exhibit large sensitivities to an effective width parameter of the rainbow-ladder interaction kernel, 
            a feature which has previously been observed in Ref.~\cite{Maris:2002yu}.

\begin{figure*}[tbp]
            \begin{center}
            \includegraphics[scale=0.7]{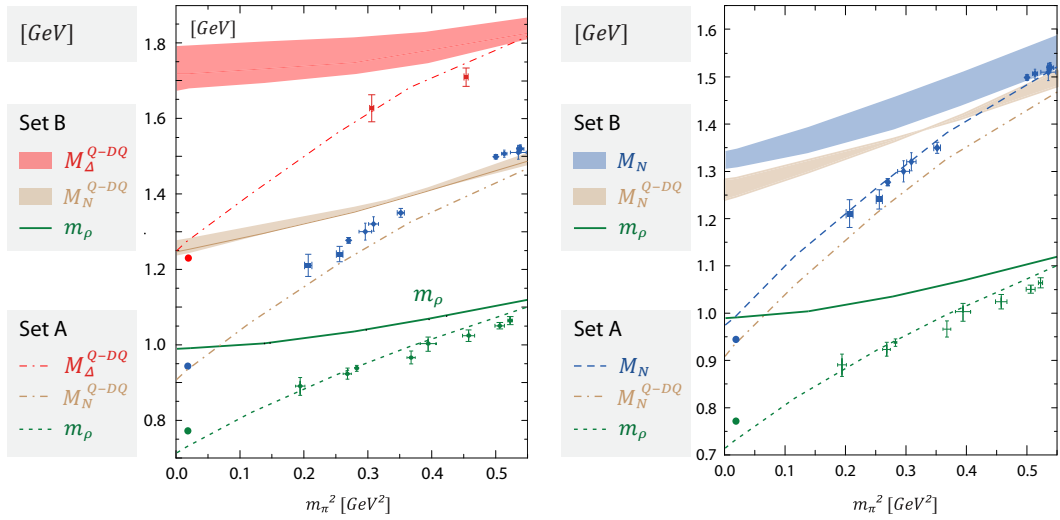}
            \caption{ (right panel adapted from Ref.\protect \cite{Eichmann:2009qa})
                                                    Evolution with $m_{\pi}^2$ of $m_\rho$, $M_N$ and $M_\Delta$
                                     compared to lattice data
                                     The left panel shows the quark-diquark model results for nucleon and $\Delta$ masses
                                     \protect \cite{Eichmann:2008ef,Nicmorus:2008vb};                                                                
                                     the right panel compares the quark-diquark and Faddeev results for the nucleon mass.
                                     Dashed and dashed-dotted lines correspond to setup A ({\it i.e.}~$f_\pi$ is fixed as input);
                                     the solid line for $m_\rho$ and the bands for $M_N$ and $M_\Delta$ are the results of setup B
				     (advertent inflation of hadronic observables to leave room for pionic corrections),
                                     where the variation w.r.t.\ to the interaction width parameter is explicitly taken into account
				     to provide error estimates.
                                     Dots denote the experimental values.} \label{fig:nucleon-mass}
            \end{center}
            \end{figure*}

                The results for nucleon and
                $\Delta$ masses 
		employing the quark-diquark bound state equation (Fig.~\ref{fig:qdqbse}) are shown in the left panel of Fig.\,\ref{fig:nucleon-mass}
                together with a selection of lattice results.
                The corresponding abscissa values $m_\pi^2$, as well as $m_\rho$, are
            	obtained from the pseudoscalar and vector-meson Bethe-Salpeter equations in one consistent calculation.
                         The findings are qualitatively similar to those for $m_\rho$: setup A, where the coupling strength
             is adjusted to the experimental value of $f_\pi$, agrees with the lattice data.
             This behaviour can be understood in light of a recent study of corrections beyond rainbow-ladder truncation
             which suggests a near cancellation in the $\rho$-meson of pionic effects and non-resonant corrections from the quark-gluon vertex  \cite{Fischer:2009jm}.
             Setup B provides a description of a quark core which on purpose overestimates the experimental values (see the discussion below) while it approaches
             the lattice results at larger quark masses.


\section{Baryons:\\ covariant Faddeev equation} \label{sec:faddeev}

            A formalism to treat the three-body bound-state problem
            through the analogue of the two-body Bethe-Salpeter equation
            was formulated in Refs. \cite{Taylor:1966zza,Boehm:1976ya}, see \cite{Loring:2001kv} for an overview.
            The corresponding covariant Faddeev equation describes the baryon as a bound state of three spin-$\nicefrac{1}{2}$ valence quarks (cf. Fig.~\ref{fig:faddeev}):
            \begin{equation} \label{faddeeveq}
                \Psi = \widetilde{K}_\text{(3)}\,\Psi\,, \qquad \widetilde{K}_{(3)} = \widetilde{K}_\text{(3)}^\text{irr} + \sum_{a=1}^3 \widetilde{K}^{(a)}_{(2)}\,,
            \end{equation}
            where the interaction kernel $\widetilde{K}_{(3)}$
            comprises a three-quark irreducible contribution and the sum of permuted two-quark kernels
            whose quark-antiquark analogues appear in a meson Bethe-Salpeter equation.
            The subscript $a$ denotes the respective accompanying spectator quark.

            The QCD Green functions once more provide a way to embed
            this equation in a consistent quantum-field theoretical setup.
            Its dynamical ingredients can then be treated in perfect correspondence with studies of quark and meson properties.
            A solution of the equation relies upon knowledge of the dressed quark propagator and the three-quark kernel. A numerically solvable form
	    of this equation is based on 
            the specification and decomposition of the Poincar\'e-covariant baryon amplitude.

            Recently a fully Poincar\'e-covariant computation of the nucleon's Faddeev amplitude was performed in such a unified treatment
	    \cite{Eichmann:2009qa,Eichmann:2009fb}.
            The assumption of dominant quark-quark correlations motivates the omission of the three-body irreducible contribution from the 
	    full three-quark kernel in Eq.~\eqref{faddeeveq}.
            The numerical solution of the Faddeev equation is then performed after
            truncating the interaction kernel to a dressed gluon-ladder exchange between any two quarks,
            thereby making a direct comparison with corresponding meson studies meaningful. Further insight into the structure of baryons is 
	    provided by contrasting the results to those of 
            investigations of baryons in the quark-diquark model.

             The resulting nucleon mass $M_N$ and its evolution with $m_\pi^2$ is plotted in the right panel of Fig.~\ref{fig:nucleon-mass}.
             A comparison to the consistently obtained quark-diquark model result exhibits a discrepancy of only $\sim 5\%$.
             This surprising and reassuring result indicates that a description of the nucleon as a superposition of scalar and axial-vector
             diquark correlations that interact with the remaining quark provides a close approximation to the consistent
             three-quark nucleon amplitude.

             By construction, the rainbow-ladder truncation omits all mesonic dressings of baryons. As pionic effects are certainly important  for all hadrons
	     this leads to the idea of changing the effective interaction such to advertently  inflate the $\rho$ mass (Set B in Fig.~\ref{fig:nucleon-mass})
	     and thus  to leave room
	     for pionic corrections at small pion masses. It can be seen that such a treatment effects also the baryon masses in a way which provides room for further 
	     attraction by pions.


\section{Nucleon Electromagnetic\\ Form Factors}

              \begin{figure*}[t]
              \begin{center}
              \includegraphics[scale=0.8]{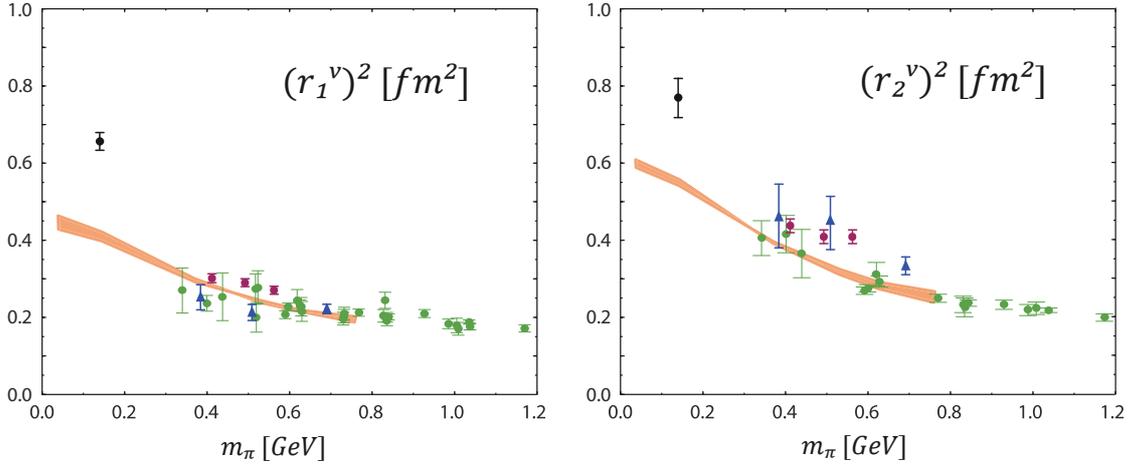}
            \caption[Radii]{(adapted from Ref.\protect \cite{Eichmann:2008kk}) 
                                       Quark-diquark model results for the squared isovector radii (corresponding to the Dirac and Pauli form factors $F_{1,2}^v=F_{1,2}^p-F_{1,2}^n$)
                                       in setup B, compared to lattice results\protect  \cite{Alexandrou:2006ru,Gockeler:2007ir}.} \label{fig:radii}
              \end{center}
              \end{figure*}

              In order to calculate the nucleon's  electromagnetic form factors in the given approach one has to
             relate the nucleon's electromagnetic current to the underlying description of the nucleon as a composite object. To this end
            the baryon must be resolved into its constituents to each of which the current can couple.
            A systematic procedure for the construction
            of a hadron-photon vertex based on electromagnetic gauge invariance \cite{Haberzettl:1997jg,Kvinikhidze:1998xn,Kvinikhidze:1999xp}
	     yields for the current operator
            \begin{equation}\label{FF:current}
                J^\mu \, =  -\conjg{\Psi}_{\!f} \left( G_0^\mu + G_0 \, K^\mu \, G_0 \right) \Psi_i\,,
            \end{equation}
            where $G_0$ denotes the product of dressed propagators and $K$ the kernel which appears in the respective bound-state equation.
            In the quark-diquark model, the incoming and outgoing baryon states are described by quark-diquark amplitudes $\Phi_i$, $\Phi_f$.
            Upon interaction with the external current, the baryon is resolved into its constituents: quark and diquark
            and the interaction between them. To each of these elements the current couples \cite{Oettel:1999gc,Eichmann:2007nn}.

              Previous nucleon form factor studies performed within the quark-diquark model \cite{Hellstern:1997pg,Bloch:1999ke,Oettel:1999gc,Oettel:2000jj,Oettel:2002wf,Bloch:2003vn}
               share some common caveats.
              First, as already indicated above
	      pionic contributions play an important role in the low-energy and small-quark mass behavior of the nucleon's electromagnetic structure.
              Such effects are not included in a quark-diquark 'core' and must be added on top of it \cite{Hecht:2002ej,Oettel:2002cw,Alkofer:2004yf,Cloet:2008re}.
              Second, access to the large-$Q^2$ region and thereby to the truly perturbative domain is so far only feasible upon implementing pole-free model propagators
              which, in turn, exhibit essential singularities at timelike infinity, cf. ref.~\cite{Alkofer:1999jf}. The problem
	      of taking into account the realistic analytic structure of the quark propagator
	       is not of fundamental concern; it merely awaits a thorough numerical treatment.
              Third, the quark-mass dependence of magnetic moments and charge radii, while emerging naturally in lattice calculations,
              is practically inaccessible in a quark-diquark model due to the unknown mass dependence of the modeled ingredients.

               The rainbow-ladder based quark-diquark approach described in Section~{\ref{sec:qdq}}
               removes the latter obstacle.
               Upon resolving the diquarks' substructure, the form factors are immediately related to
               the parameters in the effective quark-gluon coupling $\alpha(k^2)$, in particular: its quark-mass dependent coupling strength.
               Here we restrict ourselves to the quark 'core model' which represents a quark-diquark core
               that needs to be dressed by meson-cloud effects.
               A comparison of the core's static properties with lattice results is appropriate at larger quark masses;
               form factors depending on the photon momentum may be compared to experiment at $Q^2 \gtrsim 2$ GeV$^2$ where pion-cloud effects are diminished.

    The nucleon's Dirac and Pauli radii $r_1$ and $r_2$ are shown in Fig.\,\ref{fig:radii} and follow a similar pattern as the pion charge radius~\cite{Eichmann:2008ae}:
    they are weakly dependent on the width parameter of the interaction and agree with lattice data at larger quark masses where the 'quark core' becomes the baryon.
    A natural feature of a quark-diquark model is the negativity of $F_1^n(Q^2)$.
    The presence of an axial-vector $dd$ diquark correlation centers the $d$-quark in the neutron and induces $r_1^u>r_1^d$ \cite{Eichmann:2008ef}.

    Only the components transverse to the photon momentum, {\it i.e.}~those not constrained by current conservation, determine the physical form factor content.
    The respective contribution to the quark-photon vertex is known from its inhomogeneous Bethe-Salpeter solution \cite{Maris:1999bh} and includes a $\rho$-meson pole
    which amounts to $\sim 50\%$ of both pion and nucleon squared charge radii.
    The available information on those parts at larger $Q^2$, especially for the diquark-photon ingredients,
    is limited within the scope of the current approach; however they are mandatory to enable a realistic $Q^2$-evolution of the form factors
    and the proton's form factor ratio $\mu_p G_E^p/G_M^p$ \cite{Eichmann:2008ef}.
    It is noteworthy that such uncertainties will be removed upon extending the the quark-diquark approach to a three-body framework:
    in a rainbow-ladder truncation, the analogue of Eq.~\eqref{FF:current} only involves the dressed-quark photon vertex and a dressed gluon propagator, {\it i.e.},
    quantities which are known and need not be modeled.


\section{Summary and Outlook}

In Landau gauge QCD, the Green functions of quarks and gluons are sufficiently
well determined such that they can be used as a reliable input in relativistic
bound state equations. These provide then a unified approach to mesons and
baryons within a quantum-field-theoretical framework. Within the rainbow-ladder
truncation meson observables and the nucleon mass have been calculated.
These recent results mark a milestone in the effort to determine nucleon
observables in a 'first-principle' functional approach to continuum QCD.

Electromagnetic properties of the nucleon and other baryons are currently
investigated within this truncation scheme. The extensions of recent
studies of meson properties beyond rainbow-ladder \cite{Fischer:2009jm} to the
covariant Faddeev equations will improve on the now existing unified approach to
meson and baryon properties.

\newpage

\acknowledgments{RA thanks the organizers of the {\em  5-th International
Conference on Quarks and Nuclear Physics } for the invitation and
for their support. He is thankful to Yuxin Liu and Huan Chen for their
hospitality.\\
We are grateful to C.S.\ Fischer, M.\ Schwinzerl,  and R.\ Williams for useful
discussions.\\ This work was supported by  the Austrian Science Fund FWF
under Projects No. P20592-N16, P20496-N16, and Doctoral Program No. W1203,
and in part by the European Union (HadronPhysics2 project
``Study of strongly interacting matter'').}

\vspace{5mm}

\providecommand{\newblock}{}

\end{multicols}

\clearpage


\begin{thebibliography}{10}
\expandafter\ifx\csname url\endcsname\relax
  \def\url#1{{\tt #1}}\fi
\expandafter\ifx\csname urlprefix\endcsname\relax\def\urlprefix{URL }\fi
\providecommand{\eprint}[2][]{\url{#2}}

\bibitem{Fischer:2009tn}
Fischer C~S and Pawlowski J~M 2009 {\em Phys. Rev. D\/} {\bf 80} 025023
  (\textit{Preprint} \eprint{0903.2193})

\bibitem{Fischer:2008uz}
Fischer C~S, Maas A and Pawlowski J~M 2008 {\em Annals Phys.\/} {\bf 324}
  2408--2437 (\textit{Preprint} \eprint{0810.1987})

\bibitem{Alkofer:2008tt}
Alkofer R, Fischer C~S, Llanes-Estrada F~J and Schwenzer K 2009 {\em Annals
  Phys.\/} {\bf 324} 106 (\textit{Preprint} \eprint{0804.3042})

\bibitem{Alkofer:2004it}
Alkofer R, Fischer C~S and Llanes-Estrada F~J 2005 {\em Phys. Lett. B\/} {\bf
  611} 279--288 (\textit{Preprint} \eprint{hep-th/0412330})

\bibitem{Alkofer:2006fu}
Alkofer R and Greensite J 2007 {\em J. Phys. G\/} {\bf 34} S3
  (\textit{Preprint} \eprint{hep-ph/0610365})

\bibitem{Alkofer:2006gz}
Alkofer R, Fischer C~S and Llanes-Estrada F~J 2008 {\em Mod. Phys. Lett. A\/}
  {\bf 23} 1105 (\textit{Preprint} \eprint{hep-ph/0607293})


\bibitem{Eichmann:2009zx}
Eichmann G 2009 {\em {Hadron properties from QCD bound-state equations}\/}
  Ph.D. thesis University of Graz (\textit{Preprint} \eprint{0909.0703})

\bibitem{Nicmorus:2008vb}
Nicmorus D, Eichmann G, Krassnigg A and Alkofer R 2009 {\em Phys. Rev. D\/}
  {\bf 80} 054028 (\textit{Preprint} \eprint{0812.1665})

\bibitem{Eichmann:2008kk}
Eichmann G, Alkofer R, Krassnigg A and Nicmorus D 2008 {\em PoS\/} {\bf
  Confinement8} 077 (\textit{Preprint} \eprint{0812.3183})

\bibitem{Nicmorus:2008eh}
Nicmorus D, Eichmann G, Krassnigg A and Alkofer R 2008 {\em PoS\/} {\bf
  Confinement8} 052 (\textit{Preprint} \eprint{0812.2966})

\bibitem{Eichmann:2008ef}
Eichmann G, Cloet I~C, Alkofer R, Krassnigg A and Roberts C~D 2009 {\em Phys.
  Rev. C\/} {\bf 79} 012202(R) (\textit{Preprint} \eprint{0810.1222})

\bibitem{Fischer:2009jm}
Williams R, these proceedings;
Fischer C~S and Williams R 2009 {\em Phys. Rev. Lett.\/} {\bf 103} 122001
  (\textit{Preprint} \eprint{0905.2291})

\bibitem{Eichmann:2009qa}
Eichmann G, Alkofer R, Krassnigg A and Nicmorus D 2009  (\textit{Preprint}
  \eprint{0912.2246})

\bibitem{Krassnigg:2009zh}
Krassnigg A 2009 {\em Phys. Rev. D\/} {\bf 80} 114010 (\textit{Preprint}
  \eprint{0909.4016})

\bibitem{Maris:1997hd}
Maris P, Roberts C~D and Tandy P~C 1998 {\em Phys. Lett. B\/} {\bf 420}
  267--273 (\textit{Preprint} \eprint{nucl-th/9707003})

\bibitem{Holl:2004fr}
Holl A, Krassnigg A and Roberts C~D 2004 {\em Phys. Rev. C\/} {\bf 70}
  042203(R) (\textit{Preprint} \eprint{nucl-th/0406030})

\bibitem{Maris:1999nt}
Maris P and Tandy P~C 1999 {\em Phys. Rev. C\/} {\bf 60} 055214
  (\textit{Preprint} \eprint{nucl-th/9905056})

\bibitem{Eichmann:2008ae}
Eichmann G, Alkofer R, Cloet I~C, Krassnigg A and Roberts C~D 2008 {\em Phys.
  Rev. C\/} {\bf 77} 042202(R) (\textit{Preprint} \eprint{0802.1948})

\bibitem{Hess:1998sd}
Hess M, Karsch F, Laermann E and Wetzorke I 1998 {\em Phys. Rev. D\/} {\bf 58}
  111502 (\textit{Preprint} \eprint{hep-lat/9804023})

\bibitem{Wetzorke:2000ez}
Wetzorke I and Karsch F 2000  (\textit{Preprint} \eprint{hep-lat/0008008})

\bibitem{Orginos:2005vr}
Orginos K 2006 {\em PoS\/} {\bf LAT2005} 054 (\textit{Preprint}
  \eprint{hep-lat/0510082})

\bibitem{Liu:2006zi}
Liu Z and DeGrand T 2006 {\em PoS\/} {\bf LAT2006} 116 (\textit{Preprint}
  \eprint{hep-lat/0609038})

\bibitem{Alexandrou:2006cq}
Alexandrou C, de~Forcrand P and Lucini B 2006 {\em Phys. Rev. Lett.\/} {\bf 97}
  222002 (\textit{Preprint} \eprint{hep-lat/0609004})

\bibitem{Babich:2007ah}
Babich R {\em et~al.\/} 2007 {\em Phys. Rev. D\/} {\bf 76} 074021
  (\textit{Preprint} \eprint{hep-lat/0701023})

\bibitem{Maris:2002yu}
Maris P 2002 {\em Few-Body Syst.\/} {\bf 32} 41--52 (\textit{Preprint}
  \eprint{nucl-th/0204020})

\bibitem{Maris:2004bp}
Maris P 2004 {\em Few-Body Syst.\/} {\bf 35} 117--127 (\textit{Preprint}
  \eprint{nucl-th/0409008})

\bibitem{Anselmino:1992vg}
Anselmino M, Predazzi E, Ekelin S, Fredriksson S and Lichtenberg D~B 1993 {\em
  Rev. Mod. Phys.\/} {\bf 65} 1199--1234

\bibitem{Jaffe:2004ph}
Jaffe R~L 2005 {\em Phys. Rept.\/} {\bf 409} 1--45 (\textit{Preprint}
  \eprint{hep-ph/0409065})

\bibitem{Alkofer:2005ug}
Alkofer R, Kloker M, Krassnigg A and Wagenbrunn R~F 2006 {\em Phys. Rev.
  Lett.\/} {\bf 96} 022001 (\textit{Preprint} \eprint{hep-ph/0510028})

\bibitem{Hellstern:1997pg}
Hellstern G, Alkofer R, Oettel M and Reinhardt H 1997 {\em Nucl. Phys. A\/}
  {\bf 627} 679--709 (\textit{Preprint} \eprint{hep-ph/9705267})

\bibitem{Oettel:2000jj}
Oettel M, Alkofer R and von Smekal L 2000 {\em Eur. Phys. J. A\/} {\bf 8}
  553--566 (\textit{Preprint} \eprint{nucl-th/0006082})

\bibitem{Alkofer:2004yf}
Alkofer R, Holl A, Kloker M, Krassnigg A and Roberts C~D 2005 {\em Few-Body
  Syst.\/} {\bf 37} 1--31 (\textit{Preprint} \eprint{nucl-th/0412046})

\bibitem{Holl:2005zi}
H\"oll A, Alkofer R, Kloker M, Krassnigg A, Roberts C~D and Wright S~V 2005
  {\em Nucl. Phys. A\/} {\bf 755} 298--302 (\textit{Preprint}
  \eprint{nucl-th/0501033})

\bibitem{Eichmann:2007nn}
Eichmann G, Krassnigg A, Schwinzerl M and Alkofer R 2008 {\em Annals Phys.\/}
  {\bf 323} 2505--2553 (\textit{Preprint} \eprint{0712.2666})

\bibitem{Taylor:1966zza}
Taylor J~G 1966 {\em Phys. Rev.\/} {\bf 150} 1321--1330

\bibitem{Boehm:1976ya}
Boehm M and Meyer R~F 1979 {\em Annals Phys.\/} {\bf 120} 360--384

\bibitem{Loring:2001kv}
Loering U, Kretzschmar K, Metsch B~C and Petry H~R 2001 {\em Eur. Phys. J. A\/}
  {\bf 10} 309--346 (\textit{Preprint} \eprint{hep-ph/0103287})

\bibitem{Eichmann:2009fb}
Eichmann G, Alkofer R, Krassnigg A and Nicmorus D 2009  (\textit{Preprint}
  \eprint{0912.2876})

\bibitem{Alexandrou:2006ru}
Alexandrou C, Koutsou G, Negele J~W and Tsapalis A 2006 {\em Phys. Rev. D\/}
  {\bf 74} 034508 (\textit{Preprint} \eprint{hep-lat/0605017})

\bibitem{Gockeler:2007ir}
Gockeler M {\em et~al.\/} 2007  (\textit{Preprint} \eprint{arXiv:0709.3370
  [hep-lat]})

\bibitem{Haberzettl:1997jg}
Haberzettl H 1997 {\em Phys. Rev. C\/} {\bf 56} 2041--2058 (\textit{Preprint}
  \eprint{nucl-th/9704057})

\bibitem{Kvinikhidze:1998xn}
Kvinikhidze A~N and Blankleider B 1999 {\em Phys. Rev. C\/} {\bf 60} 044003
  (\textit{Preprint} \eprint{nucl-th/9901001})

\bibitem{Kvinikhidze:1999xp}
Kvinikhidze A~N and Blankleider B 1999 {\em Phys. Rev. C\/} {\bf 60} 044004
  (\textit{Preprint} \eprint{nucl-th/9901002})

\bibitem{Oettel:1999gc}
Oettel M, Pichowsky M and von Smekal L 2000 {\em Eur. Phys. J. A\/} {\bf 8}
  251--281 (\textit{Preprint} \eprint{nucl-th/9909082})

\bibitem{Bloch:1999ke}
Bloch J~C~R, Roberts C~D, Schmidt S~M, Bender A and Frank M~R 1999 {\em Phys.
  Rev. C\/} {\bf 60} 062201 (\textit{Preprint} \eprint{nucl-th/9907120})

\bibitem{Oettel:2002wf}
Oettel M and Alkofer R 2003 {\em Eur. Phys. J. A\/} {\bf 16} 95--109
  (\textit{Preprint} \eprint{hep-ph/0204178})

\bibitem{Bloch:2003vn}
Bloch J~C~R, Krassnigg A and Roberts C~D 2003 {\em Few-Body Syst.\/} {\bf 33}
  219--232 (\textit{Preprint} \eprint{nucl-th/0306059})

\bibitem{Hecht:2002ej}
Hecht M~B {\em et~al.\/} 2002 {\em Phys. Rev. C\/} {\bf 65} 055204
  (\textit{Preprint} \eprint{nucl-th/0201084})

\bibitem{Oettel:2002cw}
Oettel M and Thomas A~W 2002 {\em Phys. Rev. C\/} {\bf 66} 065207
  (\textit{Preprint} \eprint{nucl-th/0203073})

\bibitem{Cloet:2008re}
Cloet I~C, Eichmann G, El-Bennich B, Klahn T and Roberts C~D 2009 {\em Few Body
  Syst.\/} {\bf 46} 1--36 (\textit{Preprint} \eprint{0812.0416})

\bibitem{Alkofer:1999jf}
Alkofer R, Ahlig S, Fischer C~S, Oettel M and Reinhardt H 2000 {\em Nucl. Phys.
  A\/} {\bf 663} 683--686 (\textit{Preprint} \eprint{hep-ph/9907563})

\bibitem{Maris:1999bh}
Maris P and Tandy P~C 2000 {\em Phys. Rev. C\/} {\bf 61} 045202
  (\textit{Preprint} \eprint{nucl-th/9910033})

\end{thebibliography}
\end{document}